\documentclass[pra,showpacs,twocolumn,aps]{revtex4-1}

\usepackage{amsmath}
\usepackage{amssymb}
\usepackage{latexsym}
\usepackage{amsfonts}
\usepackage{epsfig}
\usepackage{psfrag}
\usepackage{graphicx}

\usepackage{epsfig,color}
\usepackage{psfrag}
\usepackage[utf8]{inputenc}
\definecolor{darkgreen}{rgb}{0,0.4,0}

\newcommand{\ket}[1]{\left|#1\right>}
\newcommand{\bra}[1]{\left<#1\right|}   

\newcommand{\nn}{\nonumber\\}

\newcommand{\f}[1]{\mbox{\boldmath$#1$}}

\newcommand{\bea}{\begin{eqnarray}}
\newcommand{\ea}{\end{eqnarray}}
\newcommand{\eea}{\end{eqnarray}}
\newcommand{\ord}{\,{\cal O}}
\newcommand{\li}{\,\widehat{\cal L}}
\newcommand{\tr}{\,{\rm Tr}}
\newcommand{\vc}[1]{\mathbf{#1}}

\begin{document}

\title{Quasi-particle approach for general lattice Hamiltonians} 

\author{Patrick Navez, Friedemann Queisser, 
and Ralf Sch\"utzhold}

\email{ralf.schuetzhold@uni-due.de}

\affiliation{
Fakult\"at f\"ur Physik, Universit\"at Duisburg-Essen, 
Lotharstrasse 1, 47057 Duisburg, Germany}

\date{\today}

\begin{abstract}
In many condensed-matter systems, it is very useful to introduce a 
quasi-particle approach, which is based on some sort of linearization around a 
suitable background state.
In order to be a systematic and controlled approximation, this linearization 
should be justified by an expansion into powers of some small control 
parameter. 
Here, we present a method for general lattice Hamiltonians with large 
coordination numbers $Z\gg1$, which is based on an expansion into powers of 
$1/Z$. 
In order to demonstrate the generality of our method, we apply it to various 
spin systems, 
as well as the Bose and Fermi Hubbard model.  
\end{abstract}


\maketitle

\section{Introduction}

Apart from a few exceptions (mostly in one spatial dimension), the majority of 
Hamiltonians describing interactions in condensed matter and other fields is 
not exactly solvable.
Thus, it is often very useful to introduce a suitable quasi-particle approach 
-- i.e., to consider linearized excitations (quasi-particles) around a given 
background (e.g., the ground state).  
However, in order to avoid neglecting contributions which are of equal 
magnitude as the terms one keeps (or even larger), such a linearization should 
be based on the expansion into powers of some small control parameter.  

As an example, Bose-Einstein condensates of dilute atomic gases are often 
described in terms of a mean-field expansion where the full field operator is 
split up via 
\bea
\hat\Psi(t,\f{r})=\psi_{\rm c}(t,\f{r})+\delta\hat\Psi(t,\f{r})
\ea
into the condensate wave-function $\psi_{\rm c}(t,\f{r})$ and the remaining 
(linearized) quantum or thermal fluctuations $\delta\hat\Psi(t,\f{r})$.
The above split can be justified if a large number $N\gg1$ of atoms share the 
same wave-function $\psi_{\rm c}(t,\f{r})$, since it scales with $\sqrt{N}$ 
in this case \cite{GA97,LL80}.

As another example, spin systems are often described using the large-$S$ 
expansion (spin-wave theory) \cite{Yosida01}.
This approximation can be motivated via the Holstein-Primakoff transformation 
which maps the spin operators to bosonic creation and annihilation operators 
$\hat a^\dagger$ and $\hat a$ 
\bea
\hat S_+
=
\hat a^\dagger\sqrt{2S-\hat a^\dagger\hat a}
=
\hat a^\dagger\sqrt{2S}
+\ord(1/\sqrt{S})
\ea
and $\hat S_-=\hat S_+^\dagger$ as well as $\hat S_z=S-\hat a^\dagger\hat a$.
For large spins $S$, the system is then approximated by non-interacting 
bosons. 
The reliability of the $1/S$-expansion for $S=1/2$ systems is another 
question, see Section~\ref{Heisenberg Model} below.

In the following, we present a general formalism for the introduction of 
quasi-particles for lattice Hamiltonians which is based on an expansion in 
powers of $1/Z$, where $Z$ is the coordination number of the lattice. 
The paper is organized as follows. Section \ref{Basic Formalism} provides 
the key elements  of the $1/Z$ method. In section \ref{Factorization}, we derive 
the formalism for quasi-particle excitation. We illustrate the usefulness of these concepts 
for calculating the excitation spectrum and the phase transition boundaries
for various known model descriptions: the Heisenberg model in section  \ref{Heisenberg Model}, 
the Ising model in section \ref{Ising Model}, the Bose-Hubbard model in section \ref{Bose-Hubbard Model} and 
the Fermi-Hubbard in section \ref{Fermi-Hubbard Model}.
The last section \ref{Conclusions} is devoted to the conclusions.

\section{Basic Formalism}\label{Basic Formalism}

We consider a general lattice Hamiltonian
\bea
\hat H=\frac1Z\sum_{\mu\nu}\hat H_{\mu \nu}+\sum_\mu\hat H_\mu
\,,
\ea
consisting of on-site terms $\hat H_\mu$ and two-site coupling or tunneling 
terms $\hat H_{\mu \nu}$ (with $\hat H_{\mu\mu}=0$) where $\mu$ and $\nu$ 
label the lattices sites. 
For simplicity, we assume that the $\hat H_\mu$ have all the same form 
(and similarly the $\hat H_{\mu \nu}$), i.e., we have (discrete) translational 
and rotational invariance of the underlying lattice.
Finally, the coordination number $Z$ counts the number of tunneling or 
interaction neighbors $\mu$ of a given lattice site $\nu$ and we consider the 
limit $Z\gg1$. 
The time-evolution of the density matrix $\hat\rho$ of the total lattice is 
given by the von Neumann equation 
\bea
\label{Liouville}
i\partial_t\hat\rho 
=
\left[\hat H,\hat\rho\right] 
= 
\frac1Z\sum_{\mu\nu}\li_{\mu \nu}\hat\rho 
+
\sum_\mu\li_\mu\hat\rho
\,,
\ea
with the Liouville super-operators 
$\li_{\mu\nu}\hat\rho=[\hat H_{\mu\nu},\hat\rho]$
and 
$\li_\mu\hat\rho=[\hat H_{\mu},\hat\rho]$.
Now we introduce the reduced density matrices $\hat\rho_\mu$
(or $\hat\rho_{\mu\nu}$) of a given lattice site $\mu$ (or $\mu,\nu$) 
by tracing over all sites except $\mu$ (or $\mu,\nu$)
\bea
\label{reduced-density-matrices}
\hat\rho_\mu 
=
\tr_{\not\mu}\{\hat\rho\}
\,,\;
\hat\rho_{\mu\nu}
=
\tr_{\not\mu\not\nu}\{\hat\rho\}
\,,\;{\rm etc.}
\ea
Note that  $\hat\rho_\mu$ is a linear operator acting on the Hilbert space 
${\cal H}_\mu$ of one lattice site $\mu$ while $\hat\rho_{\mu\nu}$ 
is a linear operator acting on ${\cal H}_\mu\otimes{\cal H}_\nu$ etc. 
Now we may split up those matrices into correlated and uncorrelated parts 
\bea
\label{correlated-parts}
\hat\rho_{\mu\nu}
=
\hat\rho_{\mu\nu}^{\rm corr}+\hat\rho_{\mu}\hat\rho_{\nu}
\,,
\ea
and 
$\hat\rho_{\mu\nu\lambda}=\hat\rho_{\mu\nu\lambda}^{\rm corr}+
\hat\rho_{\mu\nu}^{\rm corr}\hat\rho_{\lambda}+
\hat\rho_{\mu\lambda}^{\rm corr}\hat\rho_{\nu}+
\hat\rho_{\nu\lambda}^{\rm corr}\hat\rho_{\mu}+
\hat\rho_{\mu}\hat\rho_{\nu}\hat\rho_{\lambda}$ 
etc. 
From Eq.~(\ref{Liouville}), we get the equation of motion for the reduced 
density matrix $\hat\rho_\mu$ 
\bea
\label{one-site}
i\partial_t\hat\rho_{\mu}
=
\frac{1}{Z}
\sum_{\kappa\neq\mu}\tr_{\kappa}\left\{
\li^S_{\mu \kappa}
(\hat\rho^{\rm corr}_{\mu \kappa}+\hat\rho_\mu \hat\rho_\kappa)\right\}
+
\li_\mu\hat\rho_{\mu}
\,,
\eea
where $\li_{\mu \nu}^S=\li_{\mu \nu}+\li_{\nu \mu}$. 

The above equation is still exact -- but in order to solve it and to obtain 
$\hat\rho_{\mu}$, we would need the correlations 
$\hat\rho^{\rm corr}_{\mu\kappa}$.
Similarly, the equation for the two-point correlations 
$\hat\rho^{\rm corr}_{\mu\nu}$ contains the three-point correlator 
$\hat\rho^{\rm corr}_{\mu\nu\kappa}$ and so on.
However, for large coordination numbers $Z\gg1$, one can solve this set of 
equations approximately and iteratively. 
If we start in a separable (i.e., uncorrelated) initial state, the 
correlations satisfy -- at least for a finite period of time -- the following 
hierarchy \cite{NS10,QKNS13}:
\bea
\label{hierarchy}
\hat\rho_{\mu} = \ord\left(Z^0\right)
\,,\;
\hat\rho^{\rm corr}_{\mu\nu} = \ord\left(1/Z\right)
\,,\;
\hat\rho^{\rm corr}_{\mu\nu\kappa} = \ord\left(1/Z^2\right)
\,,
\ea
and so on.  
This allows us to obtain the approximate equation for the on-site density 
matrix
\bea
\label{one-site-approx}
i\partial_t\hat\rho_{\mu}
=
\frac{1}{Z}
\sum_{\kappa\neq\mu}\tr_{\kappa}\left\{
\li^S_{\mu \kappa}
\hat\rho_\mu \hat\rho_\kappa\right\}
+
\li_\mu\hat\rho_{\mu}
+\ord(1/Z)
\,.
\eea
The zeroth-order solution $\hat\rho^0_\mu$ of this equation corresponds to the 
local (e.g., Gutzwiller) mean-field ansatz.
Inserting this solution and performing an analogous procedure for two sites, 
we get 
\bea
\label{two-sites-approx}
i \partial_t \hat\rho^{\rm corr}_{\mu \nu}
&=&
\li_\mu\hat\rho^{\rm corr}_{\mu\nu}
+
\frac1Z\li_{\mu\nu}\hat\rho^0_\mu\hat\rho^0_\nu
-\frac{\hat\rho^0_{\mu}}{Z}
\tr_{\mu}
\left\{\li^S_{\mu\nu}\hat\rho^0_\mu\hat\rho^0_\nu\right\}
\nn
&&+
\frac1Z
\sum_{\kappa\not}
\tr_{\kappa}
\left\{
\li^S_{\mu \kappa}
(\hat\rho^{\rm corr}_{\mu\nu}\hat\rho^0_{\kappa}+
\hat\rho^{\rm corr}_{\nu\kappa}\hat\rho^0_{\mu})
\right\}
\nn
&&
+(\mu\leftrightarrow\nu)
+\ord(1/Z^2)
\,.
\eea
This equation determines the quantum fluctuations around the background 
solution $\hat\rho^0_\mu$ and will be the starting point for our 
quasi-particle picture.

\section{Factorization}\label{Factorization}

As it turns out, Eq.~(\ref{two-sites-approx}) can be simplified considerably 
by effectively factorizing it. 
To this end, let us introduce the propagator $\hat W_\mu^\nu(t,t_0)$ from an 
initial lattice site $\nu$ at time $t_0$ to a final lattice site $\mu$ at time 
$t$ which corresponds to the mapping 
\bea
\label{mapping}
\hat A_\nu(t_0)\,\to\,\hat A_\nu(t)
=
\sum\limits_\mu\tr_\mu\left\{\hat W_\mu^\nu(t,t_0)\hat A_\mu(t_0)\right\}
\,.
\ea
Here $\hat W_\mu^\nu(t,t_0)$ is a linear operator acting on 
${\cal H}_\mu\otimes{\cal H}_\nu$ where ${\cal H}_\mu$ and ${\cal H}_\nu$
are two different Hilbert spaces (initial and final lattice sites) 
even for $\mu=\nu$.

To achieve consistency for $t=t_0$, we impose the initial condition
\bea
\bra{n_\alpha,n_\mu} {\hat W}_{\mu}^{\alpha}(t,t) \ket{m_\alpha,m_\mu}
=
\delta_\mu^\alpha\,\delta_{n_\mu}^{n_\alpha}\,\delta_{m_\mu}^{m_\alpha}
\,,
\ea
where the $\ket{n_\mu}$ form a complete basis of ${\cal H}_\mu$ 
and similarly the $\ket{n_\alpha}$ for ${\cal H}_\alpha$.  
If we now postulate the following effective equation of motion for the 
propagator 
\begin{eqnarray}
\label{effective}
i\partial_t{\hat W}_{\mu}^{\alpha}(t,t_0)
=
\li_\mu{\hat W}_{\mu}^{\alpha}(t,t_0)
+
\nn
+
\frac{1}{Z}
\sum_{\kappa}\tr_{\kappa}\left\{\li^S_{\mu\kappa}
\left(
\hat\rho^{0}_\kappa{\hat W}_{\mu}^{\alpha}(t,t_0)
+
{\hat W}_{\kappa}^{\alpha}(t,t_0)\hat\rho^{0}_\mu
\right)
\right\}
,\;
\end{eqnarray}
and insert it into the factorization ansatz 
\begin{eqnarray}
\label{ansatz}
\hat\rho^{\rm corr}_{\mu \nu}(t)
= 
\sum_{\alpha\beta}\tr_{\alpha\beta}
\left\{
{\hat W}_{\mu}^{\alpha}(t,t_0)
{\hat W}_{\nu}^{\beta}(t,t_0)
\hat\rho^{\rm corr}_{\alpha\beta}(t_0)
\right\}
+
\nn
+\int\limits_{t_0}^t dt' 
\sum_{\alpha\beta}\tr_{\alpha\beta}
\left\{
{\hat W}_{\mu}^{\alpha}(t,t')
{\hat W}_{\nu}^{\beta}(t,t')
\hat Q_{\alpha\beta}(t')
\right\}
,\;
\end{eqnarray}
where $\hat Q_{\alpha\beta}$ denotes the sources term in 
Eq.~(\ref{two-sites-approx})
\begin{eqnarray}
\hat Q_{\alpha\beta}
&=&
%
i\,\frac{\hat\rho^0_{\alpha}}{Z}
\tr_{\alpha}
\left\{
\li_{\alpha\beta}^S\hat\rho_{\alpha}^{0}\hat\rho_{\beta}^{0}
\right\}
+i\,\frac{\hat\rho^0_{\beta}}{Z}
\tr_{\beta}
\left\{
\li_{\alpha\beta}^S\hat\rho_{\alpha}^{0}\hat\rho_{\beta}^{0}
\right\}
\nn
&&
-\frac{i}{Z}\li_{\alpha\beta}^S\hat\rho_{\alpha}^{0}\hat\rho_{\beta}^{0}
\,,
\end{eqnarray}
we find that we obtain the solution of Eq.~(\ref{two-sites-approx}) to first 
order in $1/Z$. 
Ergo, instead of solving Eq.~(\ref{two-sites-approx}) directly, we may solve 
the simpler equation (\ref{effective}) instead and recover the correlations 
$\hat\rho^{\rm corr}_{\mu \nu}(t)$ as a bilinear combination (\ref{ansatz}).

Interestingly, the effective equation (\ref{effective}) for fixed $\alpha$
(the initial lattice site) is formally equivalent to a direct linearization of 
the on-site equation (\ref{one-site-approx}).
To see that, let us set 
$\hat\rho_{\mu}=\hat\rho^{0}_\mu + \delta\hat\rho_{\mu}$ and linearize around 
the stationary solution $\hat\rho^{0}_\mu$.
The resulting equation 
\begin{eqnarray}
\label{linearize}
i\partial_t\delta\hat\rho_\mu
=
\li_\mu\delta\hat\rho_\mu
+
\frac{1}{Z}
\sum_{\kappa}\tr_{\kappa}\left\{\li^S_{\mu\kappa}
\left(\hat\rho^{0}_\kappa\delta\hat\rho_\mu
+\delta\hat\rho_\kappa\hat\rho^{0}_\mu\right)
\right\}
\nonumber \\
\end{eqnarray}
is formally equivalent to (\ref{effective}).  
Accordingly, the solution of this equation is given by (\ref{mapping}) 
\bea
\delta\hat\rho_\mu(t)
=
\sum\limits_\nu\tr_\nu \left\{\hat W_\mu^\nu(t,t_0)\delta\hat\rho_\nu(t_0)\right\}
\,,
\nonumber
\ea
which confirms that $\hat W_\mu^\nu(t,t_0)$ is the propagator. 

\section{Heisenberg Model}\label{Heisenberg Model}

The Heisenberg model is the first basic lattice model to which the $1/Z$ approach applies. The 
Hamiltonian is 
\bea
\label{Heisenberg}
\hat H=-\frac{J}{Z}\sum_{\mu\nu} T_{\mu\nu}
\hat{\f{S}}_\mu\cdot\hat{\f{S}}_\nu
\ea
Here $\hat{\f{S}}_\mu=(\hat S_\mu^x,\hat S_\mu^y,\hat S_\mu^z)$ are spin 
operators at the lattice site $\mu$ satisfying the usual $SU(2)$ commutation  
relations 
$[\hat S_\mu^a,\hat S_\nu^b]=i\delta_{\mu\nu}\epsilon^{abc}\hat S_\mu^c$.  
$J$ denotes the coupling strength. 
The lattice structure is encoded in the adjacency matrix $T_{\mu\nu}$ 
which equals unity if $\mu$ and $\nu$ are tunnelling neighbours 
(i.e., if a particle hops from $\mu$ to $\nu$) and zero otherwise. 
For simplicity, we restrict ourselves to spin $S=1/2$ systems, 
but our results can be generalized easily.

\subsection{Ferromagnetic case $J>0$}

The Heisenberg model~(\ref{Heisenberg}) is invariant under internal (spin) 
rotations, but its ground state breaks this symmetry spontaneously, i.e., 
we have a set of degenerate ground states.  
For positive $J$, any state where all the spins point into the same 
direction is an exact ground state. 
For simplicity, let us take the spin-up state:
\bea
\hat\rho_\mu^0=\ket{\uparrow}_\mu\!\bra{\uparrow}
\ea
Introducing the time-propagated spin operator 
\bea
\hat{\f{S}}_\nu(t)
=
\sum\limits_\mu\tr_\mu
\left\{\hat W_\mu^\nu(t,t_0)\hat{\f{S}}_\mu(t_0)\right\}
\,,
\ea
according to (\ref{mapping}), we obtain the following closed operatorial 
equation of motion:
\bea 
\label{eom-Heisenberg}
i\partial_t\hat S_\mu^z&=&0
\,,
\\
(i\partial_t \mp J) \hat S^\pm_{\mu} 
&=&
\mp\sum_{\nu}\frac{J T_{\mu \nu}}{Z}\hat S^\pm_{\nu} 
\,,
\eea
where we have used the notation 
$\hat S_{\mu}^\pm=\hat  S_\mu^x\pm i\hat S_\mu^y$. 
Assuming (discrete) translational symmetry of our lattice, we may 
Fourier transform these equations 
\bea 
\label{eom-Heisenberg-k}
\left(\omega \mp J[1-T_\mathbf{k}]\right)\hat S^\pm_{\mathbf{k},\omega}=0 
\,,
\eea
where $T_\mathbf{k}$ is the Fourier transform of the adjacency matrix 
$T_{\mu\nu}$ with the normalization $T_\mathbf{k=0}=1$. 
From the above equation, we may directly read off the spectrum 
\bea
\omega_{\mathbf{k}}=\pm J(1-T_\mathbf{k})
\,.
\eea
As expected, the magnons in the ferromagnetic phase are gapless 
(Goldstone modes) and their spectrum scales as 
$\omega_{\mathbf{k}}\sim{\mathbf{k}}^2$ for small $\mathbf{k}$. 
Note that the same spectrum can be derived via the $1/S$-expansion \cite{Yosida01}
mentioned in the Introduction.
However, for $S=1/2$, the $1/Z$-expansion appears better justified 
than the $1/S$-expansion.  

The positive 
frequency associated to the  excitation operator  $\hat S^-_{\mathbf{k},\omega}$
means that it costs energy to flip one spin up.
On the contrary, $\hat S^+_{\mathbf{k},\omega}$ has a negative energy which means 
that adding a spin up to a spin polarized magnet liberates energy. This process is not possible 
because already all the spin are up. Therefore, such a release of energy  
is possible provided excitations are created in pair: a creation of positive energy 
followed by a creation of negative energy so that the net energy is positive.
A switch in the sign of $J$ changes the sign 
of the frequencies and announces the change to the antiferromagnetism phase. In this case, 
the system becomes instable because energy is released by changing a spin up into a spin down.    

\subsection{Anti-ferromagnetic case $J<0$}

For negative $J=-J_A$, the ground state(s) cannot be determined exactly in general. 
Nevertheless, under certain conditions, we can get a good approximation.
In order to avoid problems with frustration, we assume that we have a 
bi-partite lattice, i.e., that we can divide the lattice into two sublattices 
${\cal A}$ and ${\cal B}$ such that for each site $\mu\in{\cal A}$ all the 
adjacent neighbouring sites $\nu$ belong to ${\cal B}$ and {\it vice versa}.
Then, for large $Z$, we may approximate the anti-ferromagnetic ground state 
by the N\'eel state \cite{Yosida01}
\bea
\hat \rho^{0}_{\mu}=
\left\{\begin{array}{ll}
|\downarrow \rangle_\mu \langle \downarrow| & \mu \in {\cal A} \\
|\uparrow \rangle_\mu \langle \uparrow|  & \mu \in {\cal B}
\end{array}\right.
\,.
\eea
A similar reasoning as in the ferromagnetic case leads to the following 
operatorial equations  
\bea 
\label{eom-Heisenberg-anti}
i\partial_t\hat S_\mu^z&=&0
\,,
\\
(i\partial_t \pm J_A) \hat S^\pm_{\mu} 
&=&
\mp\sum_{\nu}\frac{J_A T_{\mu \nu}}{Z}\hat S^\pm_{\nu} 
\quad{\rm if}\quad
\mu \in {\cal A}
\,,
\\
(i\partial_t \mp J_A) \hat S^\pm_{\mu} 
&=&
\pm\sum_{\nu}\frac{J_A T_{\mu \nu}}{Z}\hat S^\pm_{\nu} 
\quad{\rm if}\quad
\mu \in {\cal B}
\,.
\eea
After a Fourier transform, we deduce the following spectrum for the magnons
in the anti-ferromagnetic case
\bea
\omega_{\mathbf{k}}=
\pm J_A\sqrt{1-T^2_\mathbf{k}}
\,.
\eea
As expected, the spectrum is still gapless but scales as 
$\omega_{\mathbf{k}}\sim|\mathbf{k}|$ for small $\mathbf{k}$. 
Again, the same spectrum can be derived via the $1/S$-expansion \cite{Yosida01}

Also the frequencies go to zero at the antiferromagnetism-ferromagnetism transition. 
At the difference of ferromagnetism, it costs energy 
to change the spin up from spin down in sublattice ${\cal A}$ and the other way around 
for sublattice ${\cal B}$. But this process reverses
once the frequencies go to zero at the antiferromagnetism-ferromagnetism transition.

\section{Quantum Ising Model}\label{Ising Model}

As our second example, let us inmvestigate the quantum Ising model in a 
tranverse field with the Hamiltonian 
\bea
\hat H=-\frac{J}{Z}\sum_{\mu\nu} T_{\mu\nu}
\hat{S}_{\mu}^z \hat{S}_{\nu}^z - B\sum_{\mu}\hat{S}_{\mu}^x 
\,.
\ea
This model displays a quantum transition from the paramagnetic phase 
(where the transversal magnetic field $B$ dominates) to the ferromagnetic 
or anti-ferromagnetic phase, which breaks the ${\mathbb Z}_2$ spin-flip 
symmetry. 

To lowest order in $1/Z$, we employ the following variational ansatz for 
the ground state 
\bea
E_0
= 
\sum_{\mu}\tr_{\mu}\left\{\hat H_\mu \hat\rho^{0}_\mu\right\} +
\frac1Z
\sum_{\mu\nu} \tr_{\mu \nu}\left\{
\hat H_{\mu\nu} \hat\rho^{0}_\mu \hat\rho^{0}_\nu\right\}
\ea
where the reduced density matrix is given by a pure state 
$\hat\rho^{0}_\mu = |\psi\rangle_\mu \langle \psi|$ with
$|\psi\rangle_\mu= c_{\uparrow} |\uparrow \rangle_\mu + 
c_{\downarrow} |\downarrow \rangle_\mu$.
The variational ground state energy $E_0$ per site reads 
\bea
\frac{E_0}{N}=
-\frac{J}{4}(|c_{\uparrow}|^2 -|c_{\downarrow}|^2)^2 -
\frac{B}{2}(c^*_{\uparrow}c_{\downarrow}+c_{\uparrow}c^*_{\downarrow})
\,.
\ea
Minimizing this energy, we find that the ground state always has 
$\langle \hat S_{\mu}^y\rangle_0=0$, i.e., we can choose the amplitudes 
$c_{\uparrow}$ and $c_{\downarrow}$ to be real.
To lowest order in $1/Z$, the critical point is at $J=B$.
For $J<B$, the magnetic field controls 
the orientation of the spin so that the state is paramagnetic with $\langle \hat S_{\mu}^z\rangle=0$
and $\langle \hat S_{\mu}^x\rangle=1/2$ (assuming $B>0$). 
For $J>B$, on the other hand, the ferromagnetism case is associated to a permanent magnetization 
with a non-vanishing ferromagnetic order 
parameter $\langle\hat S_{\mu}^z\rangle=\pm\sqrt{1-B^2/J^2}/2$.

In complete analogy the Heisenberg model, we may derive the effective 
quasi-particle equations
\bea 
\label{spec}
i\partial_t\hat S_{\mu}^\pm 
&=& 
\pm B\hat S_{\mu}^z
\mp \sum_{\nu}\frac{2 J T_{\mu\nu}}{Z}
\left(\hat S_{\mu}^\pm \langle \hat S_{\nu}^z\rangle_0+
\hat S_{\nu}^z\langle \hat S_{\mu}^\pm\rangle_0\right) 
\,,
\nn
i\partial_t \hat S_{\mu}^z
&=&
\frac{B}{2}
(\hat S_{\mu}^+ -\hat S_{\mu}^- )
\,.
\eea
This gives the excitation spectrum 
\bea
\omega_{\mathbf{k}}
=
\pm\sqrt{4J^2\langle\hat S_{\nu}^z\rangle^2_0+B^2-2BJT_{\mathbf{k}}
\langle\hat S_{\nu}^x\rangle_0}
\,.
\ea
For the paramagnetic phase, this simplifies to 
\bea
\label{para}
\omega_{\mathbf{k}}
=
\pm\sqrt{B^2-BJT_{\mathbf{k}}}
\,.
\ea
At the difference of the Heisenberg models, here transition occurs for $J= B$ when 
the frequencies becomes imaginary.
For comparison, the spectrum in 1D obtained from exact diagonalisation is given by 
\bea \label{ising1D}
\omega_{1D}= \pm \sqrt{B^2 - 
BJ T_\vc{k}+ J^2/4}
\ea
and is well approximated by the lowest order result Eq.(\ref{para}) for large $B$.
In 1D however transition occurs for a zero frequency at $B=J/2$. Let us note here 
that, although the $1/Z$ expansion is supposed to work very well in the limit of 
large dimensions, it gives nevertheless qualitatively good results for 1D. 
 
In the ferromagnetic case, the spectrum is
\bea
\omega= \pm \sqrt{J^2 - 
B^2 T_\vc{k}}
\ea
and corresponds in the limit of weak $B$ 
to two kink excitation $2\omega_{1D}$ of 
the exact 1D spectrum Eq.(\ref{ising1D}). 

\section{Bose-Hubbard Model}\label{Bose-Hubbard Model}

After discussing two examples for spin systems, let us apply the hierarchy 
discussed above to the Bose-Hubbard model, see also \cite{QKNS13,NS10}.
This model describes (identical) bosons hopping on a lattice with the 
tunnelling rate $J$.
If two (or more) bosons occupy the same lattice site, they repel each other 
with the interaction energy $U$.
Altogether, the Hamiltonian reads
\bea
\label{Bose-Hubbard-Hamiltonian}
\hat H
=
-\frac{J}{Z}\sum_{\mu\nu} T_{\mu\nu} \hat b^\dagger_\mu \hat b_\nu
+
\frac{U}{2}\sum_{\mu} 
\hat n_{\mu}(\hat n_{\mu}-1)
\,.
\ea
Here $\hat b^\dagger_\mu$ and $\hat b_\nu$ are creation and annihilation 
operators at the lattice sites $\mu$ and $\nu$, respectively, satisfying  
the bosonic commutation relations 
$[\hat b_\mu,\hat b_\nu^\dagger]=\delta_{\mu\nu}$ 
and
$[\hat b_\mu^\dagger,\hat b_\nu^\dagger]=[\hat b_\mu,\hat b_\nu]=0$. 
As before, the lattice structure is encoded in the adjacency matrix 
$T_{\mu\nu}$ which equals unity if $\mu$ and $\nu$ are tunnelling neighbours 
(i.e., if a particle can hop from $\mu$ to $\nu$) and zero otherwise. 
Finally, $\hat n_{\mu}=\hat b^\dagger_\mu \hat b_\mu$ is the number 
operator and we assume the Mott phase 
$\hat\rho^{0}_\mu=|1\rangle_\mu \langle 1|$ in the following. 

In complete analogy to the spin operators in the previous Sections, 
we may define effective particle and  hole operators via 
$\hat h_\mu=\ket{0}_\mu\!\bra{1}$ and 
$\hat p_\mu=\ket{1}_\mu\!\bra{2}$.
Time-propagating these operators according to (\ref{mapping}), 
we deduce the following operatorial equations \cite{QKNS13}:
\bea
\label{effective0}
i\partial_t\hat{h}_\mu
&=&
\frac{J}{Z}\sum_\nu T_{\mu\nu}
\left[\hat{h}_\nu+\sqrt{2}\,\hat{p}_\nu\right]
\,,
\nonumber\\
\left[i\partial_t-U\right]\hat{p}_\mu
&=&
-\frac{J}{Z}\sum_\nu T_{\mu\nu}
\left[2\,\hat{p}_\nu+\sqrt{2}\,\hat{h}_\nu\right]
\,.
\ea
Thus, the eigenvalues obtained from this system are 
\bea
\omega_\mathbf{k}^\pm
=
\frac{U-JT_\mathbf{k}\pm \sqrt{U^2-6 J UT_\mathbf{k}+J^2 T_\mathbf{k}^2}}{2}
\,.
\eea
They 
correspond to the ones obtained in the random phase approximation \cite{SD2005,KN11}.
Inspection of the frequencies shows that they become imaginary for $J/U= 3- 2\sqrt{2}$ 
indicating the onset of instability of the Mott state towards a superfluid state.

\section{Fermi-Hubbard Model}\label{Fermi-Hubbard Model}
As our final example, let us consider the Fermi-Hubbard model, which is 
somewhat similar to the Bose-Hubbard model, but significantly more 
complicated. 
The Hamiltonian has a quite similar form as above 
\bea
\label{Fermi-Hubbard-Hamiltonian}
\hat H
=
-\frac{J}{Z}\sum_{\mu\nu,s}T_{\mu\nu}\hat{c}_{\mu,s}^\dagger\hat{c}_{\nu,s}
+U\sum_{\mu}\hat{n}_\mu^\uparrow\hat{n}_\mu^\downarrow
\,,
\ea
but now the $\hat{c}_{\mu,s}^\dagger$ and $\hat{c}_{\nu,s}$ are fermionic 
creation and annihilation operators and we have two different fermion species 
labeled by their spin $s=\uparrow$ and $s=\downarrow$. 
Depending on the values of $J$ and $U$ as well as the filling factors 
$\langle\hat{n}_\mu^\uparrow\rangle$ and 
$\langle\hat{n}_\mu^\downarrow\rangle$, 
there are various phases which all require a different ansatz for 
$\hat\rho_\mu^0$.



Let us consider the case of large and positive $U$ with small fillings 
$\langle\hat{n}_\mu^\uparrow\rangle\ll1$ and 
$\langle\hat{n}_\mu^\downarrow\rangle\ll1$ 
and where 
the temperature $T$ satisfies $U\gg T \gg J $.
In this case, double occupancy $\ket{\uparrow\downarrow}_\mu=
\hat c_{\mu,\uparrow}^\dagger \hat c_{\mu,\downarrow}^\dagger\ket{0}_\mu$ 
can be neglected and any possible spin 
ordering is washed out thermally. We may then use the following ansatz 
\bea
\hat \rho^{0}_{\mu}= 
p_0|0 \rangle_\mu \langle 0| +
p_\downarrow |\downarrow \rangle_\mu \langle \downarrow| +
p_\uparrow |\uparrow \rangle_\mu \langle \uparrow|  
\,,
\eea
where $p_0+p_\downarrow+p_\uparrow=1$ with
$\langle\hat{n}_\mu^\uparrow\rangle=p_\uparrow$ and 
$\langle\hat{n}_\mu^\downarrow\rangle=p_\downarrow$.

This ansatz provides a solution of the lowest (zeroth) order in $1/Z$
and we find that the quasi-particle modes decouple into two branches 
for the two spin degrees of freedom. 
Let us analyze the spin-up excitation 
(the results for the other spin-down branch are analogous). 
The relevant perturbations part of the density matrix read 
\bea
\delta \hat \rho_{\mu}
&=&
u_{\mu,\uparrow}|\uparrow\rangle_\mu \langle 0|+
v_{\mu,\uparrow}|\uparrow \downarrow \rangle_\mu \langle \downarrow|
\nn
&=&
u_{\mu,\uparrow}\hat h_{\mu,\uparrow}^\dagger + 
v_{\mu,\uparrow}\hat p^\dagger_{\mu,\uparrow}
\,,
\eea
where $\hat h_{\mu,\uparrow}=\hat c_{\mu,\uparrow} (1-\hat n_{\mu,\downarrow})$
and $\hat p_{\mu,\uparrow}=\hat c_{\mu,\uparrow}\hat n_{\mu,\downarrow}$ are 
effective particle and hole operators respectively.
Using this ansatz in Eq.~(\ref{linearize}) for the 
Hamiltonian~(\ref{Fermi-Hubbard-Hamiltonian}),  
we obtain the following set of equations 
\bea
i\partial_t u_{\mu,\uparrow}
&=&
-\frac{J}{Z}\sum_\kappa T_{\mu \kappa}
(p_0+p_\uparrow)(u_{\kappa,\uparrow}+v_{\kappa,\uparrow})
\nn
(i\partial_t-U)v_{\mu,\uparrow}
&=&
-\frac{J}{Z}\sum_\kappa T_{\mu \kappa}
p_\downarrow(u_{\kappa,\uparrow}+v_{\kappa,\uparrow})
\,.
\eea
After a Fourier transform, we obtain the following Hubbard expression \cite{H63,H64a,H64b}
for the eigen-frequencies
\bea
\omega_{\mathbf{k}}^\uparrow
&=&
\frac{U-JT_{\mathbf{k}}}{2}\pm 
\frac{1}{2}\sqrt{(U+JT_{\mathbf{k}})^2 -4UJT_{\mathbf{k}}p_\downarrow}
\,.
\eea

\section{Conclusions}
\label{Conclusions}

Via a controlled expansion into inverse powers of the coordination number 
$1/Z$ leading to a hierarchy of correlations \cite{QKNS13,NS10}, we establish a simple and 
elegant way to obtain quasi-particle operators and spectra.
We demonstrated the general applicability of this method by means of several 
examples such as spin systems, the Bose and Fermi-Hubbard model.  
Even though some of the results can also be obtained by other means --
such as the $1/S$-expansion -- the $1/Z$-expansion developed 
here is in several cases (such as for $S=1/2$) better justified than the 
$1/S$-expansion, for example. 

\section*{Acknowledgements} 

The authors acknowledge valuable discussions with M. Vojta, K. Krutitsky and N. ten Brinke.  
%
This work was supported by the DFG (SFB-TR12).

\end{document}